\documentclass{article}
\pdfoutput=1
\usepackage{graphicx}
\usepackage{geometry}
\begin{document}

\title{Crossover from dynamical percolation class to directed percolation class on a two dimensional lattice} 
\author{M. Ali Saif\\
Department of Physics\\
University of Amran,
Amran,Yemen.\\
masali73@gmail.com}

\maketitle
\begin{abstract}
We study the crossover phenomena from the dynamical percolation class (DyP) to the directed percolation class (DP) in the model of diseases spreading, Susceptible-Infected-Refractory-Susceptible (SIRS) on a two-dimensional lattice. In this model, agents of three species S, I, and R on a lattice react as follows: $S+I\rightarrow I+I$ with probability $\lambda$, $I\rightarrow R$ after infection time $\tau_I$ and $R\rightarrow I$ after recovery time $\tau_R$. Depending on the value of the parameter $\tau_R$, the SIRS model can be reduced to the following two well-known special cases. On the one hand, when $\tau_R \rightarrow 0$, the SIRS model reduces to the SIS model. On the other hand, when $\tau_R \rightarrow \infty$ the model reduces to SIR model. It is known that, whereas the SIS model belongs to the DP universality class, the SIR model belongs to the DyP universality class. We can deduce from the model dynamics that, SIRS will behave as an SIS model for any finite values of $\tau_R$. SIRS will behave as SIR only when $\tau_R=\infty$. Using Monte Carlo simulations we show that as far as the $\tau_R$ is finite the SIRS belongs to the DP university class. We also study the phase diagram and analyze the scaling behavior of this model along the critical line. By numerical simulation and analytical argument, we find that the crossover from DyP to DP is described by the crossover exponent $1/\phi=0.67(2)$.

\end{abstract}

\section{Introduction}
Crossover phenomena are well-known for the equilibrium phase transitions \cite{pf,ah}. From the point of view of the renormalization group theory, crossover from one class to another class occurs if more than one fixed point is embedded in the critical surface in which each fixed point corresponds to a different class. This crossover is generally interpreted by the competition of those fixed points. Through that, if there is a system characterized by irrelevant scaling fields and one relevant scaling field, then any change in the value of the relevant scaling field will cause the movement of the system from one fixed point to another fixed point. Crossover phenomena are useful for understanding the critical phenomena in phase transitions \cite{pf,ah,hen,fi}. In nonequilibrium phase transitions with absorbing states the crossover phenomena between the universality classes of those kinds of transitions, has been observed in many models. For example the crossover from directed percolation universality class (DP) to compact directed percolation universality class (CDP) \cite{men,lub}, from parity conserving class (PC) (directed Ising (DI) class) to DP class \cite{odo}, pair contact process with diffusion (PCPD) to PC \cite{par}, DP to DI \cite{par}, PCPD to DP \cite{par,da}, mean field PCPD to the PCPD \cite{par0}, DI to DP \cite{par1}, isotropic percolation (IP) to DP \cite{zho}, DP to mean field \cite{mes}, Mean field to DP \cite{san}, Manna universality class to mean field \cite{lu}, non-DI critical behavior of the branching annihilating attracting walk (BAAW) with infinite range of attraction to DI \cite{par2} and DyP to DP \cite{dam}.

In this work, we are going to study the crossover phenomena from dynamical percolation class (DyP) to directed percolation class (DP) in the model of diseases spreading, Susceptible-Infected-Refractory-Susceptible (SIRS) on a two-dimensional lattice. In this direction, DyP universality class (general epidemic process) has been found in some models of epidemics spreading with immunization (no reinfection) such as the SIR model \cite{gras,car} or forest fire model with immune trees \cite{alb} or even in prey-predator model \cite{an,ar}. DyP class considers a generalization of DP class including the effect of immunization. Geometrical interpretation of the DyP class as an ordinary percolation system (isotropic percolation) has been introduced by Stauffer and Aharony \cite{sta}. In the other direction, the DP universality class is the most famous in the field of nonequilibrium phase transition from the active phase to the absorbing phase \cite{hen,od,hin,pa,al}. This class is robust concerning microscopic changes and many models have been found to be within this class. According to Janssen and Grassberger \cite{jan,gr}, the models which display a continuous phase transition into a unique absorbing state with a positive single component order parameter and short-range interactions without quenched disorder or additional symmetries should belong to the DP class. By any way this class has been found to be more general, systems with more than one absorbing state or long-range interaction have been found to be within the DP class (see Refs. \cite{hen,od,hin,pa} and references therein).

\section{Model and methods}
The model of epidemics spreading SIRS on the networks can be described as follows \cite{kup,ali}: consider a population of $N$ agents residing on the sites of a two-dimensional square lattice (of size $L\times L$), in which each agent is connected to $k$ of its first neighbors. The agents can exist in one state of the following three states, susceptible $(S)$, infected $(I)$, and refractory $(R)$. The interaction between the agents on the lattice occurs as follows: the agents in state $I$ on the network can infect any one of their neighbors which are in state $S$ with probability $\lambda$ at each time step. The agents in state $I$ pass to the state $R$ after an infection time $\tau_I$. The agents in state $R$ cross to the state $S$ after a recovery time $\tau_R$. During the $R$ phase, the agents are immune and do not infect. The dynamics of the model are summarized as 
\begin{eqnarray}
S+I\stackrel{\lambda}{\rightarrow} I+I ,
I\stackrel{\tau_I}{\rightarrow} R ,
R\stackrel{\tau_R}{\rightarrow} I
\end{eqnarray}

The SIRS model is a general model of disease spreading which we can map into the following two extremist special cases. The first case occurs when $\tau_R\rightarrow 0$, in this case, the SIRS model reduces to the SIS model. On the other hand second case occurs when $\tau_R\rightarrow \infty$, in this case SIRS model reduces to the SIR model. The SIS model is the model of disease spreading without immunization, however SIR model is the model of disease spreading with a perfect immunization (vanishing reinfection probability). It is well-known that, whereas the SIS model belongs to the DP universality class \cite{gra}, the SIR model belongs to the DyP universality class \cite{gras,sta}. In this work, we are interested in studying the crossover behavior from the DyP universality class to the DP universality class of this model on a two-dimensional lattice, when the value of $\tau_R$ is changed from $0$ to $\infty$. Thus the crossover control parameters here are $\lambda$ and $\tau_R$. However, for technical reasons we modify the control parameter $\tau_R$ to be $\tau=\frac{\tau_I}{\tau_R}$ and for simplicity we fix the value of $\tau_I$ to be $\tau_I=1$. 

From the model dynamics we can deduce that, initially and during the time which is $t\leq\tau_R$, the SIRS model will behave as if it is an SIR model. During this time the agents pass only from the state $S$ to $I$ followed by moving from the state $I$ to $R$, no one of the agents on the lattice can move from the state $R$ to $S$. However as the time reaches the value of $t>\tau_R$, there will be agents in the state $R$ that are ready to move from the state $R$ to $S$. This moving for the agents from the state $R$ to $S$ will alter the behavior of the model from SIR-like to SIS-like. So we can say that SIRS will behave temporarily (for a time equal to $\tau_R$) as the SIR model at the beginning of the time evolution of this model. Higher values of $\tau_R$ will put the model for a longer time in the SIR-like regime, but eventually, the SIRS will change its behavior to become similar to SIS behavior. Because of that, we can conclude that, for any finite values of $\tau_R$, soon or later SIRS will change its behavior from SIR-like to SIS-like. This model will behave exactly as SIR only when $\tau_R=\infty$ ($\tau=0$), in this case, the critical value of $\lambda$ is exactly given by $\lambda_c=0.5$ \cite{sta}. 

To determine the critical points of our model as well as the critical exponents we use the dynamical Monte Carlo method \cite{hin,gra1,jens}. In the seed simulations of this method, the time evolution of quantities such as the average number of active sites $N(T)$ and survival probability $P(t)$ exhibit a power-law behavior at criticality. Mathematically we can write this as follows:
\begin{eqnarray}
N(t)\sim t^\theta 
\end{eqnarray} 
and
\begin{eqnarray}
P(t)\sim t^{-\delta}
\end{eqnarray} 
where $\theta$ and $\delta$ are the critical exponents. Those critical exponents in $d=2$ are $\delta=0.4505$, $\theta=0.2295$ for DP class \cite{hin,mun} and $\delta=0.092$, $\theta=0.586$ for DyP class \cite{mun}.
Using Eqs. 2 and 3 we can determine the critical points and the critical exponents, however for a more accurate determination we adopt here the local slope analysis with the effective exponent which is defined as follows \cite{hin}
\begin{eqnarray}
\theta(t)=\frac{\ln[N(t)]-\ln[N(t/m)]}{\ln[m]}
\end{eqnarray}
where $m$ is a constant larger than one. Same definition we can find for the effective exponent of $\delta(t)$.

We simulate our model on a two-dimensional lattice with periodic boundary conditions. We fix the value of infection time to be $\tau_I=1$. Each agent on the lattice is connected to its first four nearest neighbors, and the system updates synchronously. In Fig. 1 we show two examples of the results which we find from the seed simulations. In our Monte Carlo simulations, we use a lattice of size $4000\times 4000$ and we take the average over $2000$ configurations. Fig. 1 shows the value of the effective exponent Eq. 4 as a function of inverse time $1/t$ for the value of $\tau=0.004$ (left) and $\tau=0.5$ (right). The corresponding critical points are $\lambda_c(0.004)=0.4755(3)$ and $\lambda_c(0.5)=0.4009(3)$. As it is clear from the figure in both cases the critical exponent $\theta$ converges to its value in the DP universality class $\theta=0.229$. We assert that, our results using the critical exponents $\theta$ or $\delta$ confirm the phase transition to be of DP universality class along the phase transition line $\lambda_c(\tau)$ except when $\tau=0$ (see Fig. 2). We have checked the validity of this conclusion up to small values of $\tau$, i. e. $\tau=0.001$ which corresponds to $\lambda_c(0.001)=0.4986(4)$. Fig. 2 shows the value of $\theta$ as a function of $1/t$ for the case when $\tau=0$ in which the critical point is exactly $\lambda_c(0)=0.5$. As it is clear from the figure the critical exponent $\theta$ approaches its value in DyP class $\theta=0.586$. 
\begin{figure}[htb]
\includegraphics[width=70mm,height=60mm]{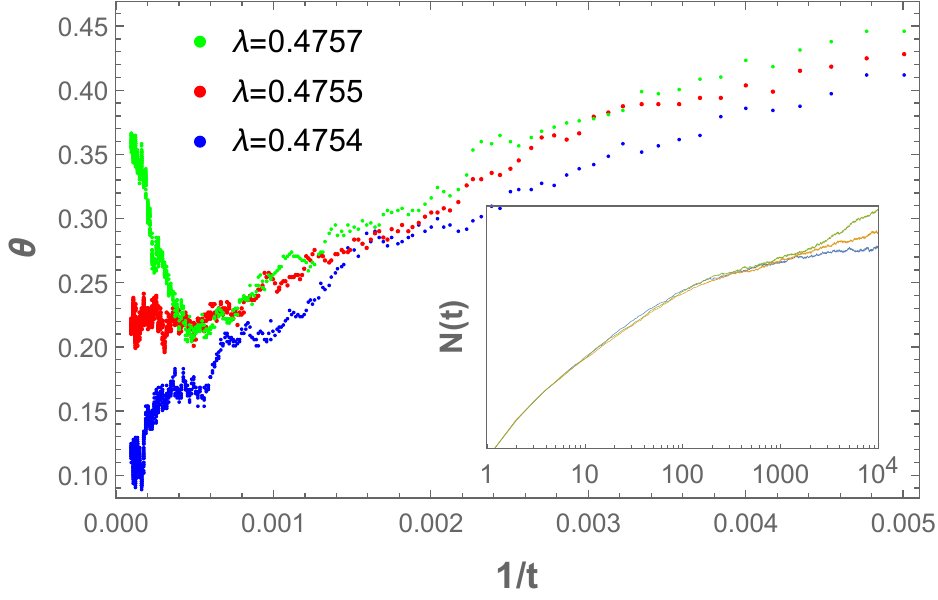}
\includegraphics[width=70mm,height=60mm]{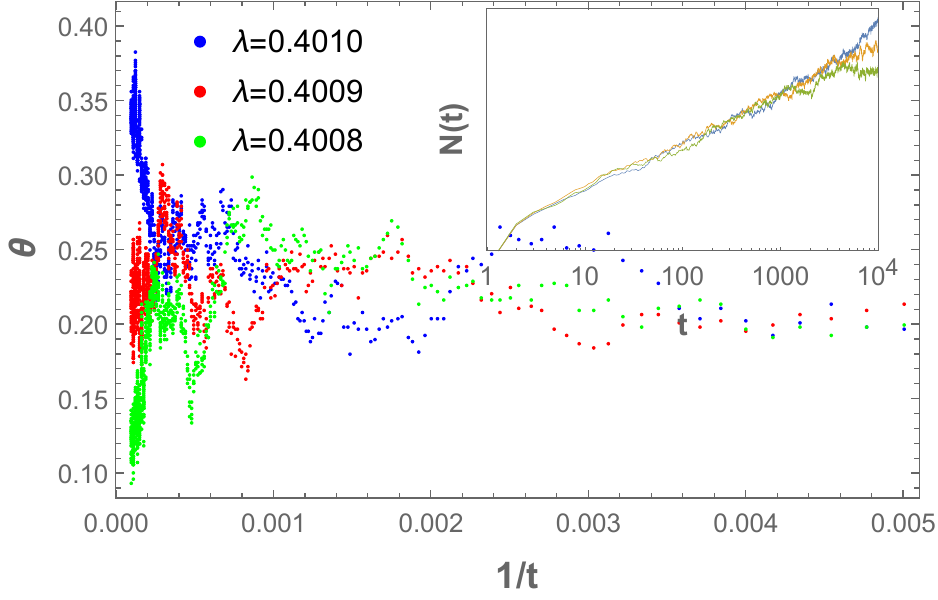}
\caption{Time-dependent behavior of the effective exponent for the average number of active sites exponent $\theta$ as a function of $1/t$ for the values of $\tau=0.004$ (left) and $\tau=0.5$ (right), for a lattice of size $4000\times 4000$. In both figures $\theta$ converges to its value in the DP class. Inset shows the number of infected agents $N(t)$ as a function of time. The data are averaged over $2000$ realization.}
 \end{figure}
\begin{figure}[htb]
\includegraphics[width=70mm,height=60mm]{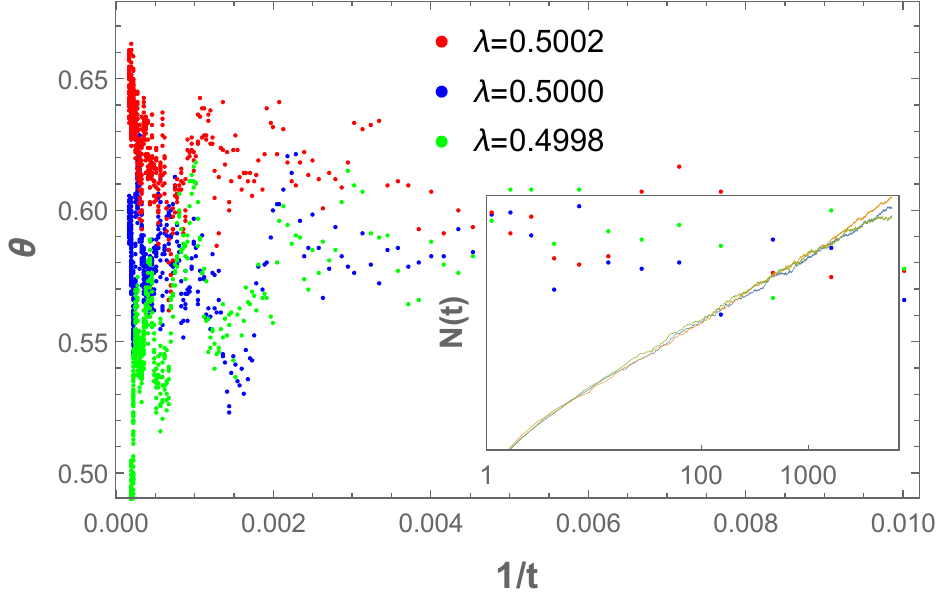}
\caption{Time-dependent behavior of the effective exponent for the average number of active sites exponent $\theta$ as a function of $1/t$ when $\tau=0$. $\theta$ converges to its value in the DyP class. Inset shows the number of infected agents $N(t)$ as a function of time.}
 \end{figure}
 
Fig. 3 shows the phase diagram of this model, in which we plot the values of the critical points $\lambda_c(\tau)$ as a function of $\tau$. The phase transition line separates the active phase (above) from the absorbing phase (below). Each point in Fig. 3 is due to the Monte Carlo simulations of lattice of size $4000\times 4000$ and, has been determined  using the effective exponent Eq. 4 as we have described previously. Figure shows clearly that, the critical line approaches the DyP class at $\tau=0$ with infinite slope (see inset of Fig. 3). The critical points we have found are summarized in Table 1.\\

\begin{figure}[htb]
\includegraphics[width=70mm,height=60mm]{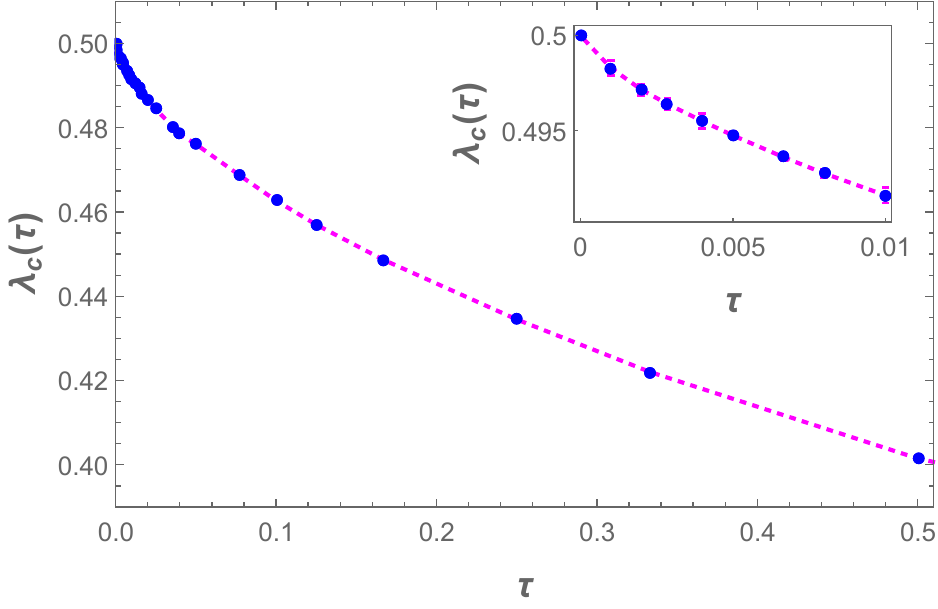}
\caption{Phase diagram of the model in the $\tau$-$\lambda$ plane. The active (absorbing) phase is above (below) the curve. The error in the critical points is smaller than the symbol size. Inset: shows the critical line beside $\tau=0$.}
 \end{figure}

\begin{tabular}{cc}
\multicolumn{2}{l}{Table 1: Critical points of the SIRS for various values of $\tau_R$.}\\
\hline
\hline
$\tau_R$ & $\lambda_c$ \\
\hline
2& 0.4015(2) \\
10& 0.4627(3)\\
50&0.4865(3)\\
100&0.4916(4)\\
200&0.4947(4)\\
350&0.4964(3)\\
500&0.4972(2)\\
1000&0.4983(4)\\
$\infty$&0.5\\
\hline
\hline
\end{tabular}\\

\section{Crossover behavior}
We are going here to study the crossover behavior of this model from the DyP class to the DP class. The control parameters of this model are $\lambda$ and $\tau$. Let us consider the case when $\tau=0$, in this case, the SIRS model reduces to the SIR model which belongs to DyP class \cite{gras}. The critical point for the system in this case is $\lambda_c=0.5$ \cite{sta}. Now from renormalization methods, we know that the phase transitions into absorbing states are characterized usually by simple scaling laws. For example, the scaling behavior for the average number of active sites $N(t)$ near the critical point is described by the function \cite{law,nm} 
\begin{eqnarray}
N(\Delta_\lambda;t)= t^\theta \psi(\Delta_\lambda t^{1/\nu_\|})
\end{eqnarray} 
where $\psi(\Delta_\lambda t^{1/\nu_\|})$ is a scaling function, $\Delta_\lambda=\lambda_c-\lambda$ denotes the distance from criticality and $\nu_\|$ is critical exponent corresponding to temporal correlation length which can be given by the relation
\begin{eqnarray}
\xi_\|\approx \left|\Delta_\lambda\right|^{-\nu_\|}
\end{eqnarray}

Now when $\tau>0$, we can modify the function Eq. 5 to include the additional control parameter $\tau$ to be
\begin{eqnarray}
N(\Delta_\lambda,\tau;t)= t^\theta \Psi(\Delta_\lambda t^{1/\nu_\|},\tau t^{1/\mu_\|})
\end{eqnarray} 
where $\mu_\|$ is the critical exponent of temporal correlation length corresponding to the parameter $\tau$. Like $\nu_\|$ this exponent can be given by the relation
\begin{eqnarray}
\xi_\|\approx \tau^{-\mu_\|}
\end{eqnarray}

Comparing Eq. 6 with Eq. 8 we can define the crossover exponent by the shape of phase boundary $\lambda_c(\tau)$ approaches the critical point $\lambda_c(0)$ as the relevant scaling field $\tau\rightarrow 0$ as follows
\begin{eqnarray}
\Delta_\lambda(\tau)\approx \tau^{1/\phi}
\end{eqnarray}
where $\Delta_\lambda(\tau)=\lambda_c(\tau)-\lambda_c(0)$ and the crossover exponent will be
\begin{eqnarray}
1/\phi=\frac{\mu_\|}{\nu_\|}
\end{eqnarray}
In Fig. 4 we show a double-logarithmic plot of Eq. 9 near the DyP critical point $\lambda_c(0)=0.5$. The points given in Fig. 4 are the same as those given in Fig. 3 when the values of $\tau<<1$. For the best estimating, the value of the crossover exponent is $1/\phi=0.67(2)$.
\begin{figure}[htb]
\includegraphics[width=70mm,height=60mm]{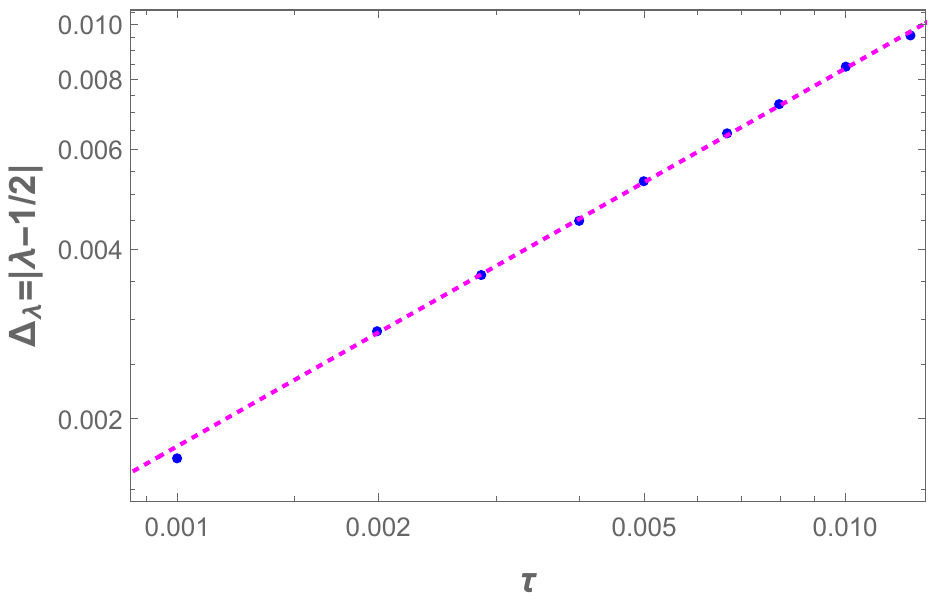}
\caption{Double-logarithmic plot of the phase transition line. The dashed line is power-law fitting on the numerical data points using Eq. 9 with the crossover exponent $1/\phi=0.67(2)$.}
\end{figure}

At this point, we are going to suggest the following method to estimate the value of crossover exponent $1/\phi$ for this model. As we have mentioned previously, initially and during the time which is proportional to $\tau_R=\tau^{-1}$, the SIRS model will behave as the SIR model. During this time the agents on the lattice pass from the state $S$ to $I$ and from the state $I$ to $R$, no one of the agents can pass from the state $R$ to $S$. After this time the first infected agents on the lattice will start departing from the state $R$ to $S$ consequently, the behavior of the system changes to be similar to the behavior of the SIS model. In Fig. 5 we manifest this behavior of this system where we plot the average value of active sites $N(t)$ (left) and the average value of survival probability $P(t)$ (right) as a function of time for values of $\tau=0, 0.004, 0.04$ and $0.5$ at the corresponding critical points $\lambda_c(\tau)$. In both figures, the upper line represents the DyP behavior ($\tau=0$), whereas the lower line corresponds to the DP class ($\tau=0.5$).  In between those two lines, the figures show how the system starts as DyP class for some time after that reaches DP class.  
\begin{figure}[htb]
\includegraphics[width=70mm,height=60mm]{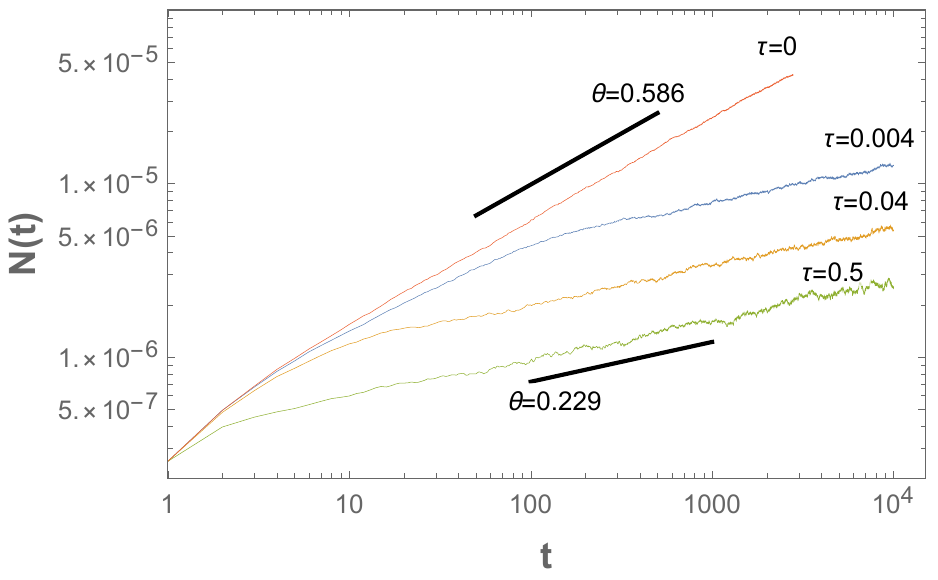}
\includegraphics[width=70mm,height=60mm]{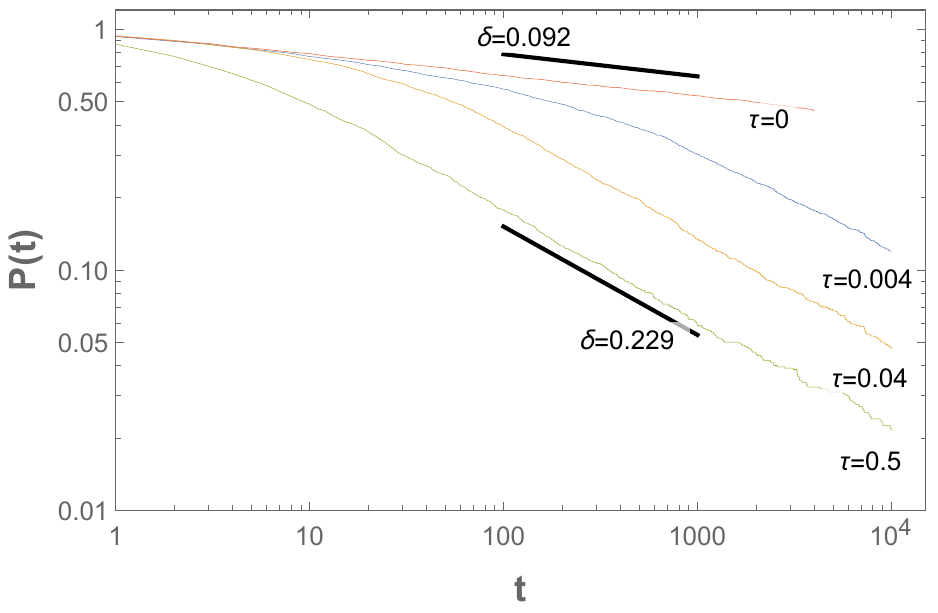}
\caption{Number of active sites $N(t)$ (left) and survival probability $P(t)$ (right) as a function of time for values of $\tau=0, 0.004, 0.04$ and $0.5$ at the corresponding values of critical points $\lambda_c(\tau)$.}
\end{figure}

Therefore we can conclude that, initially the process behaves as a critical DyP until the first refractory agents on the lattice move from the state $R$ to $S$ at a typical time $\xi_\|$. Here we assume that $\xi_\|$ scales in the same way as the typical time at which the first refractory agents on the lattice start moving from the state $R$ to $S$ \cite{dam}. Whereas the time at which the first refractory agents on the lattice start moving from the state $R$ to $S$ is proportional to $\tau^{-1}$, so we can conjecture also that $\xi_\|$ is proportional to $\tau^{-1}$
\begin{eqnarray}
\xi_\|\propto \tau^{-1} 
\end{eqnarray}  
comparing this relation with the relation in Eq. 8 we can deduce the value of $\mu_\|$ to be $\mu_\|=1.0$. 

To check the validity of this value of $\mu_\|$, we use here the scaling function of the system besides its criticality. For that, we set $\lambda=0.5$ in the scaling function Eq. 7 to get  
\begin{eqnarray}
N(0,\tau;t) t^{-\theta}=  \Psi(0,\tau t^{1/\mu_\|})
\end{eqnarray} 
Fig. 6 shows the collapse of data into a single curve using the exponent $\mu_\|=1.00(4)$ as the Eq. 12 suggests. Inset of Fig. 6 shows the average number of active sites $N(t)$ as a function of time when $\lambda=0.5$, for the values of $\tau=0.001, 0.002$ and $0.004$. This result confirms what we found previously for this value of $\mu_\|$.    

Now, using this value of the exponent $\mu_\|=1.0$ and the value of the critical exponent $\nu_\|=1.506$ for DyP class in $d=2$ \cite{mun}, and the Eq. 10, we can estimate the value of $1/\phi$ to be equal to
\begin{eqnarray}
1/\phi=\frac{\mu_\|}{\nu_\|}=\frac{1}{1.506}=0.664
\end{eqnarray}
 
this value of $1/\phi$ coincides very well with its the value that we have estimated using the Monte Carlo simulations of this model.
\begin{figure}[htb]
\includegraphics[width=70mm,height=60mm]{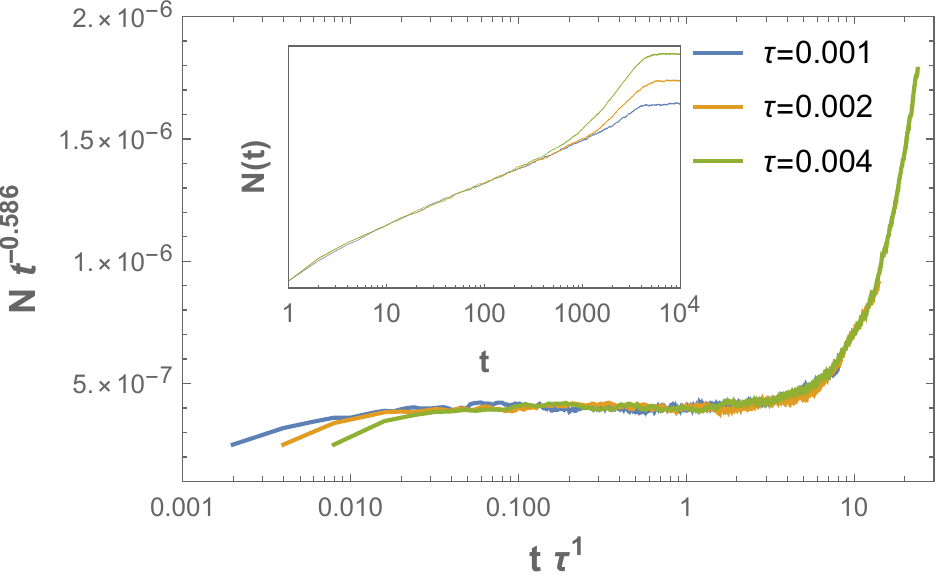}
\caption{Semilogarithmic plot of $N t^{-\theta}$ as a function of $t \tau^{\mu_\|}$ with $\mu_\|=1.0$ for $\tau=0.001, 0.002$ and $0.004$. Curves collapse well into a single curve.}
\end{figure}

Finally, we mention that crossover from DyP class to DP class has been observed in the model of epidemic spreading with immunization and mutations in two dimensions \cite{dam}. However, the value of crossover exponent $1/\phi$ has been found in Ref. \cite{dam} is different from the value we find here. For comparison Table 2 gives the values of $1/\phi$ and $\mu_\|$ for both models.\\
 
\begin{tabular}{|c|c|c|}
\multicolumn{3}{l}{Table 2: Comparison of values $1/\phi$ and $\mu_\|$.}\\
\hline
 & $1/\phi$ & $\mu_\|$\\
\hline
This work& 0.67(2) & 1.00(4)\\
\hline
Ref. \cite{dam}& 0.41(3) & 0.63(4)\\
\hline
\end{tabular}\\

Such this difference in the value of the crossover exponent $1/\phi$ has been observed in the case of the crossover from DI class to DP in one dimension \cite{par1} and the crossover between directed percolation models \cite{pa2}. According to S.- C. Park and H. Park \cite{par1,pa2} the difference in the value of the crossover exponent reflects the difference in the route that leads to the crossover from one class to another class. Using that, the crossover behavior from one fixed point to another is related to the crossover operator, which forces the system to cross from one to another. In our model, the route to crossover implies only one kind of pathogen and the reinfection occurs due to the temporary immunization\footnote{We can consider the time $\tau_R$ at which the agent is in state $R$ as the agent memory} (finite memory) of agents. In case the agents are perfectly immune (infinite memory) after the first infection, the reinfection will not occur, in which the system will behave as SIR. This leads to the conclusion that memory, is crucial in establishing the crossover in our model. In the model of Ref. \cite{dam} the agents become perfectly immune after each infection however, reinfection occurs due to the pathogen mutation. Hence in this model, the pathogen mutation establishes the crossover.

\section{Conclusion}   
Depending on the value of infection time $\tau_R$ the model of epidemics spreading SIRS can show the following two extremist special cases. In the case of $\tau_R=0$ SIRS model reduces to the SIS model which belongs to the DP universality class. On the other hand when $\tau_R=\infty$ SIRS reduces to the SIR model which belongs to the DyP universality class. In this work, we have studied the crossover phenomena for this model from the DyP universality class to the DP universality class. we have found that the SIRS with finite $\tau_R$ does belong to the DP universality class. Only the case when $\tau_R=\infty$ the model is in the DyP universality class. Here we mention that similar behavior to what is found here has been observed in the BAAW model in one dimension when the range of attraction $R$ is varied \cite{par2}. In the BAAW model only the case when $R=\infty$ is different from the DI class. Anyway, Monte Carlo simulations of our model suggest the crossover exponent from the DyP class to the DP class to be $1/\phi=0.67(2)$. We have confirmed this result with an analytical argument using the fact that SIRS will behave initially as critical SIR for time which is proportional to $\tau_R=\tau^{-1}$. This suggests the temporal correlation length $\xi_\|$ also is proportional with $\tau^{-1}$ which leads us to estimate the value of the temporal correlation exponent $\mu_\|$ to be $\mu_\|=1.0$. With this value of $\mu_\|=1.0$ we have determined the value of crossover exponent using the relation $1/\phi=\frac{\mu_\|}{\nu_\|}=0.664$ which agrees very well with the numerical estimation.

Furthermore, we found that the crossover from DyP class to DP class has been observed in the model of epidemics spreading with immunization and mutations in a two dimension \cite{dam}. However, the crossover exponent we found in this work is different from its value that has been calculated in Ref. \cite{dam}. The difference in the value of the crossover exponent for both models backs to the difference in route (operator) that leads to crossover from DyP class to DP class \cite{par1}. By inspecting the dynamics of both models we can understand the difference in the route which leads to crossover. In our model, there is only one kind of pathogen and the agents become immune for a time $\tau_R$ before they pass again to $S$ state. Hence the memory induces the crossover in our model from DyP with perfect memory to DP class with temporary memory. In the model of Ref. \cite{dam}, the agents become perfectly immune after each infection however the pathogen mutation leads to reinfection. Thus the pathogen mutation is the route to crossover from the DyP class without mutation to the DP class with mutation.

 \end{document}